\documentclass[12pt,preprint]{aastex}

\def\lea{\mathrel{<\kern-1.0em\lower0.9ex\hbox{$\sim$}}}
\def\gea{\mathrel{>\kern-1.0em\lower0.9ex\hbox{$\sim$}}}

\slugcomment{Submitted to The Astrophysical Journal Letters}

\shorttitle{ACS Globular Clusters: M54}
\shortauthors{Siegel et al.}

\begin{document}

\title{The ACS Survey of Galactic Globular Clusters\footnotemark[1]:
M54 and Young Populations in the Sagittarius Dwarf
Spheroidal Galaxy}
\footnotetext[1]{The fourth paper in the series based on observations with the NASA/ESA {\it Hubble Space Telescope},
obtained at the Space Telescope Science Institute, which is operated
by AURA, Inc., under NASA contract NAS 5-26555, under programs
GO-10775 (PI: Sarajedini).
}
\author{Michael H. Siegel\altaffilmark{2}, Aaron Dotter\altaffilmark{3}, 
Steven R. Majewski\altaffilmark{4}, Ata Sarajedini\altaffilmark{5}, 
Brian Chaboyer\altaffilmark{3}, David L. Nidever\altaffilmark{4}, Jay Anderson\altaffilmark{6},
Antonio Mar\' \i n-Franch\altaffilmark{5,7}, Alfred Rosenberg\altaffilmark{7}, 
Luigi R. Bedin\altaffilmark{8}, 
Antonio Aparicio\altaffilmark{7}, Ivan King\altaffilmark{9}, 
Giampaolo Piotto\altaffilmark{10}, I. Neill Reid\altaffilmark{8}}

\altaffiltext{2}{University of Texas, McDonald Observatory, 1 University Station, C1402, Austin TX, 78712
(siegel@astro.as.utexas.edu)}
\altaffiltext{3}{Department of Physics and Astronomy, Dartmouth College,\\ 
6127 Wilder Laboratory, Hanover, NH 03755 \\ (Aaron.L.Dotter@dartmouth.edu,
chaboyer@heather.dartmouth.edu)}
\altaffiltext{4}{Dept. of Astronomy, University of Virginia,
P.O. Box 400325, Charlottesville, VA 22904-4325 (srm4n@virginia.edu, dln5q@virginia.edu)}
\altaffiltext{5}{Department of Astronomy, University of Florida, 211 Bryant Space Science
Center, Gainesville, FL 32611 (ata@astro.ufl.edu, amarin@astro.ufl.edu)}
\altaffiltext{6}{Department of Physics and Astronomy, Rice University MS-108, Houston, TX 77005 (jay@eeyore.rice.edu)}
\altaffiltext{7}{Instituto de Astrof\'\i sica de Canarias, V\'\i a L\'{a}ctea s/n, E-38200 La Laguna, 
Spain (alf@iac.es,antapaj@iac.es,amarin@iac.es)}
\altaffiltext{8}{Space Telescope Science Institute, 3700 San Martin Drive, Baltimore MD 21218
(bedin@stsci.edu, inr@stsci.edu)}
\altaffiltext{9}{Dept. of Astronomy, Univ. of Washington, Box 351580, Seattle, WA 98195-1580
(king@astro.washington.edu)}
\altaffiltext{10}{Dipartimento di Astronomia, Universit\`{a} di Padova, 35122 Padova, Italy 
(piotto@pd.astro.it)}

\begin{abstract}
We present new {\sl Hubble Space Telescope} photometry of the massive globular cluster M54 (NGC~6715) and
the superposed core of the tidally disrupted Sagittarius (Sgr) dSph galaxy 
as part of the ACS Survey of Galactic Globular Clusters.  
Our deep ($F606W$$\sim$$26.5$), high-precision photometry yields an unprecedentedly
detailed color-magnitude diagram showing the extended blue
horizontal branch and multiple main sequences of the M54+Sgr system. 
The distance and reddening to M54 are revised using
both isochrone and main-sequence fitting 
to $(m$$-$$M)_0$$=$$17.27$ and $E(B$$-$$V)$$=$$0.15$.  Preliminary assessment 
finds the M54+Sgr field to be dominated 
by the old metal-poor populations of Sgr and the globular cluster.
Multiple turnoffs indicate the presence of 
at least two intermediate-aged star formation epochs with 4 and 6 Gyr ages
and [Fe/H]=-0.4 to -0.6. We also clearly show, for the first time, a prominent, 
$\sim$2.3 Gyr old Sgr population of near-solar abundance.
A trace population of even younger ($\sim$0.1-0.8 Gyr old), more metal-rich 
([Fe/H]$\sim0.6$) stars is also indicated.
The Sgr age-metallicity relation is consistent with a closed-box model and 
multiple (4-5) star formation bursts 
over the entire life of the satellite, including the time since 
Sgr began disrupting.
\end{abstract}


\section{Introduction}
M54 (NGC~6715) is the second most massive Galactic globular cluster and, at first blush, a 
canonical ``old halo'' cluster: ancient, metal poor and with a very extended
blue horizontal branch (HB; Harris 1996).  However, M54 has been shown (Ibata et al. 1994;
Sarajedini \& Layden 1995, hereafter ``SL95"; Majewski et al. 2003; Monaco et al. 2005a; 
Siegel et al. 2007)
to lie at the photometric center and distance of the Sagittarius (Sgr) dwarf 
spheroidal (dSph) galaxy, which is merging with the Milky Way (Ibata et al. 1994), with tidal 
arms encircling the Galaxy (e.g., Ibata et al. 2001b, Newberg et al. 2002; Majewski et al. 2003). 
This has prompted discussion of whether M54 may be the nucleus of the Sgr dSph 
(e.g., SL95; DaCosta \& Armandroff 1995; Bassino \& Muzzio 1995; Layden \& Sarajedini 2000, hereafter
``LS00") around which later star formation occurred. However, based on 
the existence of two distinct ancient, metal-poor populations (MPPs) with different 
radial profiles (SL95), and the observation that Sgr would be nucleated
even if M54 were ignored (Monaco et al. 2005a), it seems likely that M54 formed
separately from Sgr and was pulled
into the dSph center through dynamical friction. 

Photometric studies of M54 provide information on both the cluster and 
the Sgr core (SL95, Layden \& Sarajedini 1997; LS00), and, 
in combination with spectroscopic efforts, have confirmed the metal-poverty and
ancient age of both the cluster ([Fe/H]=-1.5 to -1.8, 14-15 Gyr; Brown et al. 1999, LS00) 
and Sgr's distinct MPP ([Fe/H]=-1.3, age=10-11 Gyr, LS00).  
Large surveys of
Sgr's core are dominated by intermediate stellar populations 
([Fe/H]=-0.4 to -0.7, 5-8 Gyr; SL95, LS00, Bellazzini et al. 2006a; hereafter B06),
though this is likely due to Sgr's MPPs having been
selectively stripped into tidal tails (Chou et al. 
2007).  There have also been hints of a young metal-rich population (2.5 Gyr,
-0.4 dex; SL95, LS00), 
including stars of solar-abundance (Smecker-Hane \& McWilliam 2002; Monaco et al. 2005b, 
Chou et al. 2007).
When the properties of stellar populations in and around M54 are combined with 
Sgr clusters near the core (Terzan 7, Terzan 8, Arp 2; Ibata et al. 1994) or
in the Sgr tidal stream (Pal 12, Dinescu et al. 2000; Pal 2, Majewski et al. 2004; Whiting 1, 
Carraro et al. 2006), Sgr has an age-metallicity relation
(AMR) consistent with a simple closed-box model (see Fig.\ 18 of LS00).

Because Sgr has been disrupting for 
at least 2.5-3.0 Gyr (Law et al. 2005), and likely longer, 
it is a unique laboratory for exploring star formation 
in the context of hierarchical galaxy formation. 
Not only can we potentially connect the star formation
and enrichment history of a specific satellite galaxy with relevant 
timescales and events in its interaction, but also 
ascertain in detail the populations this 
disintegrating system has donated and is still donating to the Galaxy. 
We contribute to this effort by clarifying the 
Sgr+M54 stellar populations with new {\sl HST/ACS/WFC} photometry 
from the {\it ACS} Survey of Galactic Globular Clusters 
(Sarajedini et al. 2007, hereafter Paper I).

\section{Observations and Data Reduction}

The $ACS$ survey observed 65 globulars 
in the $F606W$ ($\sim$$V$) and $F814W$ ($\sim$$I$) filters
with {\sl HST/ACS/WFC}.
PSF photometry (Anderson et al., 2007, {\it in prep}) is Vega-calibrated using
the charge-transfer efficiency corrections of Reiss \& Mack (2004), calibration
procedures in Bedin et al. (2005) and zero points of Sirianni et al. (2005).  
This is supplemented by photometry of isolated saturated stars from short exposures
salvaged by summing all associated charge --- a procedure 
previously applied by Gilliland (2004).
For M54, the observation and reduction pipeline produces twelve magnitudes of precise
photometry from nearly the
tip of the red giant branch (RGB) to several magnitudes below the main sequence turnoff (MSTO).
Our extraction of
nearly 390,000 detections in the {\sl ACS/WFC} field leaves a star-subtracted image nearly
devoid of flux. However, many of the detections do not provide
precise photometry due to source faintness, charge bleeding, cosmic rays and close neighbors.
For this initial examination of the color-magnitude diagram (CMD)
we used the trend of quality-of-fit against
magnitude to select 60,000 sources with the most star-like profiles and  
$<10$\% of the flux in their PSF aperture from other stars.

\section{Color-Magnitude Diagram Features}

The CMD of the M54 field (Fig.\ 1a) shows an
extraordinary array of features:  an extended blue and red HB, at least two red RGBs, multiple MSTOs and multiple 
MSs. Fig.\ 1b shows a Hess diagram overlayed with a schematic description of 
the features seen in the complex M54+Sgr CMD, which we now detail:

\begin{figure*}[t!]
\epsscale{1.0}
\caption{ACS photometry of the M54 field. Panel (a) shows the CMD of 60,000 stars
selected to be PSF-like
and to have less than 10\% contribution of neighbor stars to their integrated light. 
Panel (b) shows a Hess
diagram of the field with an overlayed schematic describing the various populations in 
the M54+Sgr field.  The dotted line is the Sgr MPP as defined in LS00.  Panel (d) shows the 
Hess diagram overlayed with theoretical isochrones 
describing the inferred stellar populations
while Panel (c) shows the simulated CMD described in the text.}
\end{figure*}

{\it The Old M54 Population:} 
The most prominent feature in Fig.\ 1a is the strong MS and RGB from the 
combined M54 and
Sgr MPPs (shown to have slightly different RGBs by SL95 and LS00).
The red line in Fig.\ 1b is a fiducial through the MS and RGB centers for the 
MPPs determined using techniques described in 
Rosenberg et al. (2006) that interactively fit Gaussians to the top 20\% of the 
magnitude-color distribution orthogonal to the cluster sequence in 0.4 mag wide
overlapping bins stepped every 0.04 mag 
between $F814W$=15 and 25.\footnote{The metal-poor SGB shows evidence of 
bifurcation from the overlapping MPPs of M54 and Sgr.  
However, our Gaussian fits follow the center of the dominant M54 sequence.}
The Gaussian widths provide
weights for $\chi^2$ fits of main-sequence fiducials and isochrones.
The MPPs are also reflected in
the asymptotic giant branch and lengthy 
HB running from a prominent red HB through the RR Lyrae gap to an extended blue HB.  
The latter also includes the ``blue hook" population of extremely hot HB stars 
identified by Rosenberg et al. (2004).  
The extreme HB stars are centrally concentrated in the field, consistent with membership in the 
old, metal-poor cluster.
Similar ``blue hook" stars have also been identified in the massive 
clusters $\omega$ Centauri and NGC~2808 (Moehler et al. 2002, 2004).

The narrowness of the MPP sequence allows a straight-forward measurement of
the distance and reddening to M54 using MS and isochrone fitting.  To minimize any 
confusing effect of Sgr's old MS, we fit distance and reddening both by eye and with a $\chi^2$ minimization
routine to the fiducial line shown in Fig.\ 1b, which is defined by the dominant M54 MS and RGB.
Metal-poor isochrones were taken from Dotter et al. (2007, Paper II)
with [$\alpha$/Fe]=+0.2 in agreement 
with Brown et al. (1999) and [Fe/H] allowed to vary from -2 to -1.5 (in 0.1 dex steps).
The optimal fit uses an isochrone of 13 Gyr with [Fe/H]=-1.8 at $(m-M)_0$=17.23 and
$E(B-V)$=0.17.  The Hess diagram (Fig.\ 1d) 
indicates good agreement between the theoretical isochrone (red line) and the observed MPP sequence.
Empirical MS-fitting used the fiducials defined in Paper I for the NGC~6752 and M92 clusters 
weighted heaviest near $M_{F606W}$$\sim$$+4.0$. Interpolating between the two clusters to the 
isochrone abundance of M54
([Fe/H]$=$$-1.8$) yields $E(B$$-$$V)$=0.14 and $(m$$-$$M)_0$=17.31.
The mean of the MS and isochrone measures --- $(m$$-$$M)_0$$=$17.27, $E(B$$-$$V)$=0.15\footnote{We checked
the data for indications of differential reddening using methods outlined in Paper I.  The results indicate 
minimal differential reddening.} ---
is in reasonable agreement with
the RGB tip distance of Monaco et al. (2004) and the RR Lyrae distance of LS00.

Our subsequent analysis assumes that M54 and Sgr have the same reddening and distance
because M54 appears to lie at the center of Sgr (\S1) and 
the distance uncertainty ($\sim0.05$ mag or 0.6 kpc) is large enough to mask any
small discrepancy.  A small difference in distance modulus could 
slightly alter the inferred ages of the Sgr populations, as shown in B06.

{\it The Intermediate Population(s): }
The intermediate Sgr population (``SInt"),
described in detail by
LS00 and B06 (as ``Population A"), dominates wide-field surveys of Sgr.
The SInt features in Fig.\ 1b (orange) include a prominent red clump, a redder RGB and a 
redder MS that begins to diverge from the MPP MS below $F606W=24$.
Monaco et al. (2005b) and Sbordone et al. (2006) determined the abundances of the primary intermediate 
Sgr population
to be [Fe/H]=-0.4 and [$\alpha$/Fe]=-0.2.  While B06 suggest an age for this population of 
8$\pm$1.5 Gyr, at $(m$$-$$M)_0$$=$$17.2$ their SInt age is 5-6 Gyr.
Isochrones corresponding to 
the measured abundances and younger B06 age match
the SInt RGB and the bluer MSTO emerging just beyond the MPP MSTO.
However, the CMD appears to have a broad (or perhaps two distinct) MSTOs between the old and 
``young" MSTOs, which indicates multiple bursts. 
[Fe/H]=-0.6, 6 Gyr and [Fe/H]=-0.5, 4.5 Gyr isochrones (orange lines, Fig.\ 1d) seem to best
reproduce the most apparent intermediate MSTO features.

{\it The Young Population(s): }
The bluest strong MSTO (``SYng", green line in Fig.\ 1b)
corresponds to a significantly younger population than SInt.  
Though hinted at before 
(Mateo et al. 1995; Bellazzini et al. 1999a,b; LS97; LS00), only weak constraints 
could be applied to what was previously an indistinct CMD feature.  
The superior $ACS$ photometry of the M54+Sgr core, however, clearly reveals this 
as a young metal-rich MSTO with a convective hook at $F606W$$=$$19.5$.
The best fit to SYng is a 2.3 Gyr isochrone with 
[Fe/H]=-0.1 and [$\alpha$/Fe]=-0.2 (green line, Fig.\ 1d),\footnote{A satisfactory
fit can also be obtained with a 1.75 Gyr, solar abundance isochrone with $Y$=0.33. 
At this time, however, we have no reason to suspect an enhanced He abundance in Sgr.}
similar to, but more metal-rich
than, the youngest Sgr population described in LS00.
SYng is younger than the minimum interval over which
Sgr 
has been disrupting ($\sim$2.5-3 Gyr ago, Law et al. 2005).

Finally, a sparse, bright MS can be seen 
above the other MSs in the CMD and extending blueward as a ``spray" of
stars extending above the 2.3 Gyr MSTO through the blue HB (blue features in Fig.\ 1b). While the latter
could be blue straggler stars, a very young, metal-rich population (``SVYng")
is also hinted at by the 
clump of stars below the Sgr red clump 
which are too blue to be a metal-rich or intermediate RGB
bump but faint enough to be a young red HB clump.  
The lack of a distinct MSTO or RGB associated with SVYng and the potential confusing contribution of binaries
and blue stragglers
makes further statements regarding its age/composition speculative.  However, 
the SVYng stars could represent the youngest, most metal rich M54+Sgr population.  The bright MS
extends beyond the sample isochrones at [Fe/H]=$+0.56$, 100 and 800 Myr (blue lines)
overlayed in Fig.\ 1d, suggesting protracted, recent star formation in Sgr.

\begin{figure}[h]
\epsscale{1.0}
\plotone{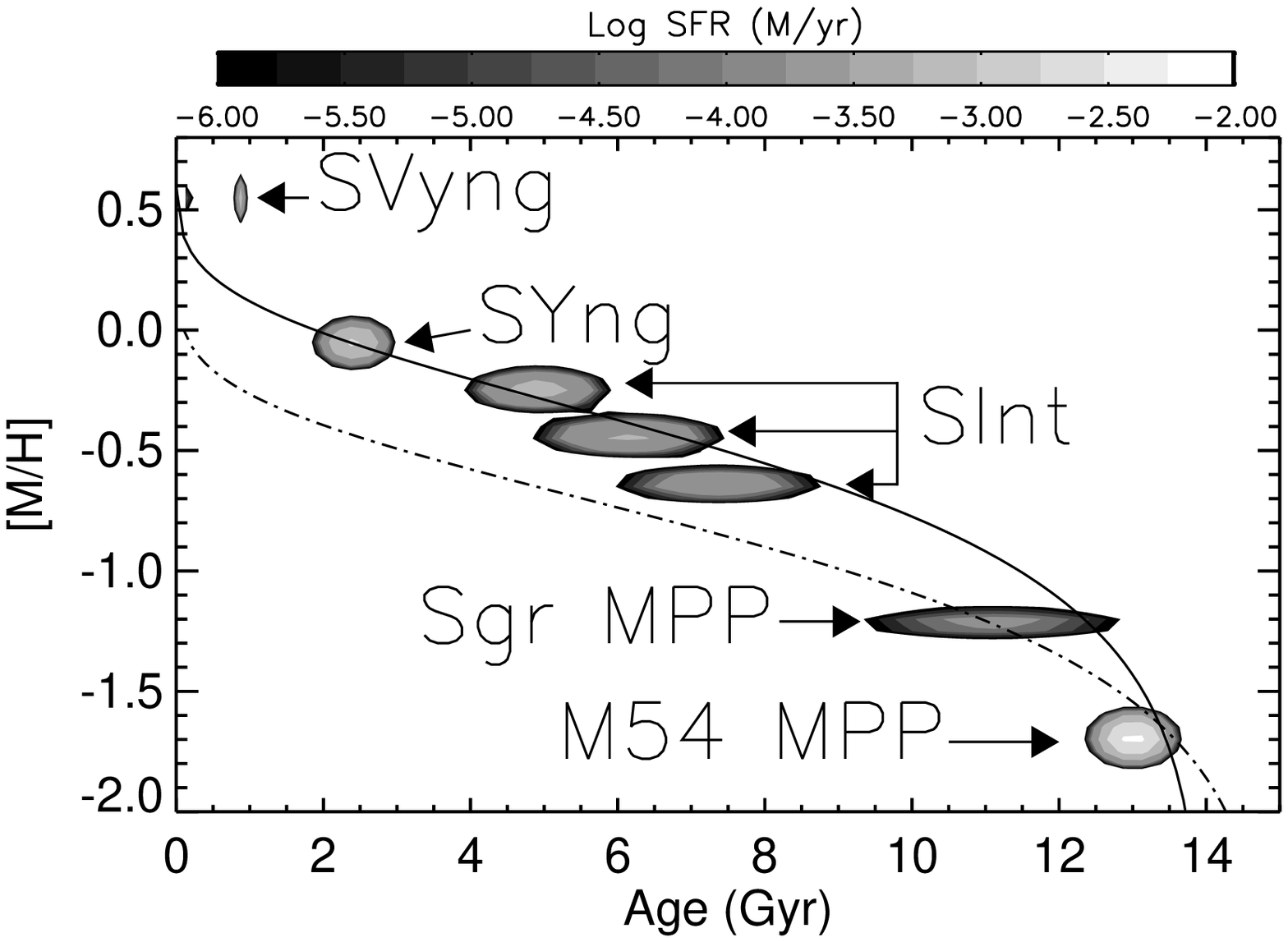}
\caption{The simulated SFH of the M54 field.  Distinct contributions are from the metal-poor M54
population (M54 MPP), Sgr's metal-poor (Sgr MPP), and Sgr's young (SYng) populations
The intermediate Sgr population (SInt) is broad and composed of
multiple bursts or continuous star formation. There appears to be
some contribution from a very young Sgr population (SVYng).
The dotted line is the AMR from LS00 using a simple closed-box model; the solid line an updated model
with faster enrichment.}
\end{figure}

\section{The Star Formation History of the M54 Field}

Untangling the multiple stellar populations of Sgr, including the contribution of binaries, can be
aided by population synthesis.
Using the isochrones fit in \S3, we reconstructed the star formation 
history (SFH) of the M54 field using the StarFISH population synthesis code 
(Harris \& Zaritsky 2001).  StarFISH uses a set of isochrones, an error
model (in this case, an analytical one) and fixed $(m-M)$, $E(B-V)$ and IMF 
to construct a library of CMD probability functions.  It then iteratively finds
which combination of synthetic CMDs reproduces the observed CMD, varying the amplitude of the 
input populations by downhill simplex until convergence.  The result
is an age-metallicity-amplitude SFH of the field.

We set the distance and reddening to the values derived in \S3 and used a Salpeter IMF.
After some initial variation, we fixed the
well-defined MPP and SYng populations while allowing
SVYng to vary in age from 0.1 to 0.9 Gyr. The SInt populations
were allowed to initially vary 
from [Fe/H]$=$$-0.3$, [$\alpha$/Fe]$=$$-0.2$ to [Fe/H]=$-1.5$, [$\alpha$/Fe]=$+0.2$ and over 2-15 Gyr.  
This window was gradually narrowed and a final fit was composed by hand to provide better age
definition and reproduction of the MSTO region


The derived M54+Sgr SFH (Fig.\ 2)\footnote{Isochrones of similar age-abundance combinations
create degenerate solutions in StarFISH. This is accounted for by ``locking" together degenerate groups of isochrones
into single CMD probability functions. The points in Figure 2 are Gaussians set to the center and range of
each locked isochrone group.} is dominated by the M54 MPP, which contributes $\sim75$\% of the simulated 
stars. Sgr contributes
a small MPP and a broad range of SInt stars.  SYng is strong and distinct while SVYng is weak and tenuous
in the CMD (Fig.\ 1c).
The populations follow a closed-box AMR model similar to that of LS00 (dotted line) but with
faster enrichment (solid line).

We supplemented the StarFISH-simulated
MSs and RGBs with synthetic HBs constructed with the He-burning tracks and modeling code from 
Paper II.  Mass distributions were constructed with an upper limit supplied by the fitted isochrones with
an average mass loss of 0.1 $M_{\odot}$  for the M54 MPP, 0.05 $M_{\odot}$ for the Sgr MPP and SInt,
and no mass loss for SYng.  The amount of mass loss for each population was set to best reproduce 
the observed HB.  All models used a mass loss dispersion of 0.05 $M_{\odot}$.  The number 
of HB stars for each population was set by the
appropriate R-ratio given the assumed He abundance.

Fig.\ 1c shows the simulated CMD.  Our relatively simple simulations recreate the salient features 
of the M54+Sgr field, including the 
broad MS, the complex MSTO, the bifurcated SGB\footnote{Our analysis of NGC~1851 (Piotto et al. 2007) 
also shows a bifurcated SGB, which we ascribe to a 1 Gyr age spread.}, the doubled RGB and 
the long blue HB. The simulated HB has a 
steeper slope than the real HB, which could be corrected if the M54 abundance 
were raised by a few 0.1 dex.

While SYng and SVYng are stronger than the SInt population(s) in our $ACS$ field, 
this does not apply to Sgr over larger scales, where other studies   
(e.g., LS00 and B06) show SInt dominating.  Surveys of Sgr's tidal arms have
shown that the stars it is contributing to the halo are, on average, even more metal-poor 
(Bellazzini et al. 2006b;
Chou et al. 2007).  Our analysis reveals the presence of recently formed stars in 
the center of Sgr, further affirming the strong metallicity gradient in the system.

\end{document}